\documentclass[12pt]{article}
\usepackage[dvips]{graphicx}

\newcommand{\ic}{\'{\i}}

\newcommand{\ala}{$\mathrm{{Al}_2O_3}$}
\newcommand{\alb}{$\mathrm{Al_{2}O_{3}:Mg}$}
\newcommand{\alc}{$\mathrm{\alpha-Al_{2}O_{3}:Mg}$}
\newcommand{\ald}{$\mathrm{Al_{2}O_{3}:Mg\sharp1}$}
\newcommand{\ale}{$\mathrm{Al_{2}O_{3}:Mg\sharp2}$}
\newcommand{\alf}{$\mathrm{\alpha-Al_{2}O_{3}}$}
\newcommand{\mg}{$\mathrm{[Mg]^{0}}$}
\newcommand{\mA}{$\mathrm{\mu A}$}
\newcommand{\cm}{$\mathrm{cm^{-3}}$}
\newcommand{\cma}{$\mathrm{cm^{2}}$}
\newcommand{\cpa}{${\it c_{\,\parallel}} $}
\newcommand{\cpe}{${\it c_{\perp}}$}
\newcommand{\lic}{$\mathrm{[Li]^{0}}$}
\newcommand{\li}{$\mathrm{Li^{+}}$}
\newcommand{\mgi}{$\mathrm{Mg^{2+}}$}
\newcommand{\ali}{$\mathrm{Al^{3+}}$}
\newcommand{\Ti}{$\mathrm{Ti^{4+}}$}
\newcommand{\Cr}{$\mathrm{Cr^{4+}}$}
\newcommand{\ohc}{$\mathrm{\Omega^{-1}cm^{-1}}$}
\newcommand{\Cp}{\it $C_{p}$}
\newcommand{\Cs}{\it $C_{s}$}
\newcommand{\Rs}{\it $R_{s}$}

\baselineskip

\begin{document}

\vspace*{-0.8cm}
\begin{flushright}
{\large cond-mat/YYYY}\\ {December-2000}\\
\end{flushright}

\makeatletter
\def\btt#1{\texttt{\@backslashchar#1}}
\DeclareRobustCommand\bblash{\btt{\@backslashchar}}
\makeatother


\begin{center}
\begin{large}
\begin{bf} SEMICONDUCTING CHARACTERISTICS OF\protect \\
 MAGNESIUM-DOPED {\alf} SINGLE CRYSTALS.\\
\end{bf}
\end{large}


\vspace{0.4cm} M. Tard{\'\i}o $^{\star}$, R. Ram{\'\i}rez, R.
Gonz{\'a}lez\\
\vspace{0.05cm} {\em  Departamento de F{\'\i}sica\\
  Escuela Polit{\'e}cnica Superior\\
Universidad Carlos III de Madrid\\ Avda. de la Universidad 30,
 28911 Legan{\'e}s (Madrid), Spain } \\
\vspace{0.05cm} Y. Chen\\
\vspace{0.06cm} {\em Division of Materials Sciences,\\ Office of
Basic Energy Sciences, SC 13, Germantown \\
MD 20874-1290, USA}\\
\vspace{0.05cm}
and\\
\vspace{0.1cm}
M.R. Kokta\\
{\em  BICRON Crystal Products, Washougal\\
Washington 96871, USA}

\end{center}

\date{\today}

\begin{abstract}
\noindent DC and AC electrical measurements were performed to
investigate the electrical conductivity of {\alc} samples with
different concentrations of {\mg} centers (Mg ions each with a
trapped hole) in the temperature interval 250-800 K. The
concentration of {\mg} centers was monitored by the optical
absorption peak at 2.56 eV. These centers were produced by
oxidation at temperatures above 1050 K. The formation rate of
{\mg} centers depends on the previous thermal history of the
sample in either reducing or oxidizing atmosphere.

At low electrical fields, DC measurements reveal blocking
contacts. At high fields, the I-V characteristic is similar to
that of a diode (corresponding to a blocking contact at one side
of the sample and an ohmic contact at the other side) connected
in series with the bulk resistance of the sample. Steady
electroluminescence is emitted at the negative electrode when  a
current in excess of $\approx$ 10 {\mA} passes through the sample,
indicating that the majority of carriers are holes.

Low voltage AC measurements show that the equivalent circuit for
the sample is the bulk resistance in series with the junction
capacitance (representing the blocking contacts) connected in
parallel with a capacitance, which represents the dielectric
constant of the sample. The values determined for the bulk
resistance in both DC and AC experiments are in good agreement.
The electrical conductivity of {\alb} crystals increases linearly
with the concentration of {\mg} centers, regardless of the amount
of other impurities also present in the crystals, and is four
times higher in the $c_{\perp}$ than in the  $c_{||}$ direction.
The conductivity is thermally activated with an activation energy
of 0.68 eV, which is independent of: 1) the {\mg} content, 2) the
crystallographic orientation, and 3) the concentration of other
impurities. These results favor the {\it small-polaron-motion}
mechanism.

\end{abstract}
\vspace{0.05cm} Key words: Electrical conductivity, $[Mg]^{0}$,
center, ${Al}_2 O_3$:$Mg$\\ \vspace{0.4cm} ${\star}$ {\em e-mail:
mtardio@fis.uc3m.es}\\



\newcommand{\figuno}{
\begin{figure}[bht]
\centering
\includegraphics*[scale=0.75]{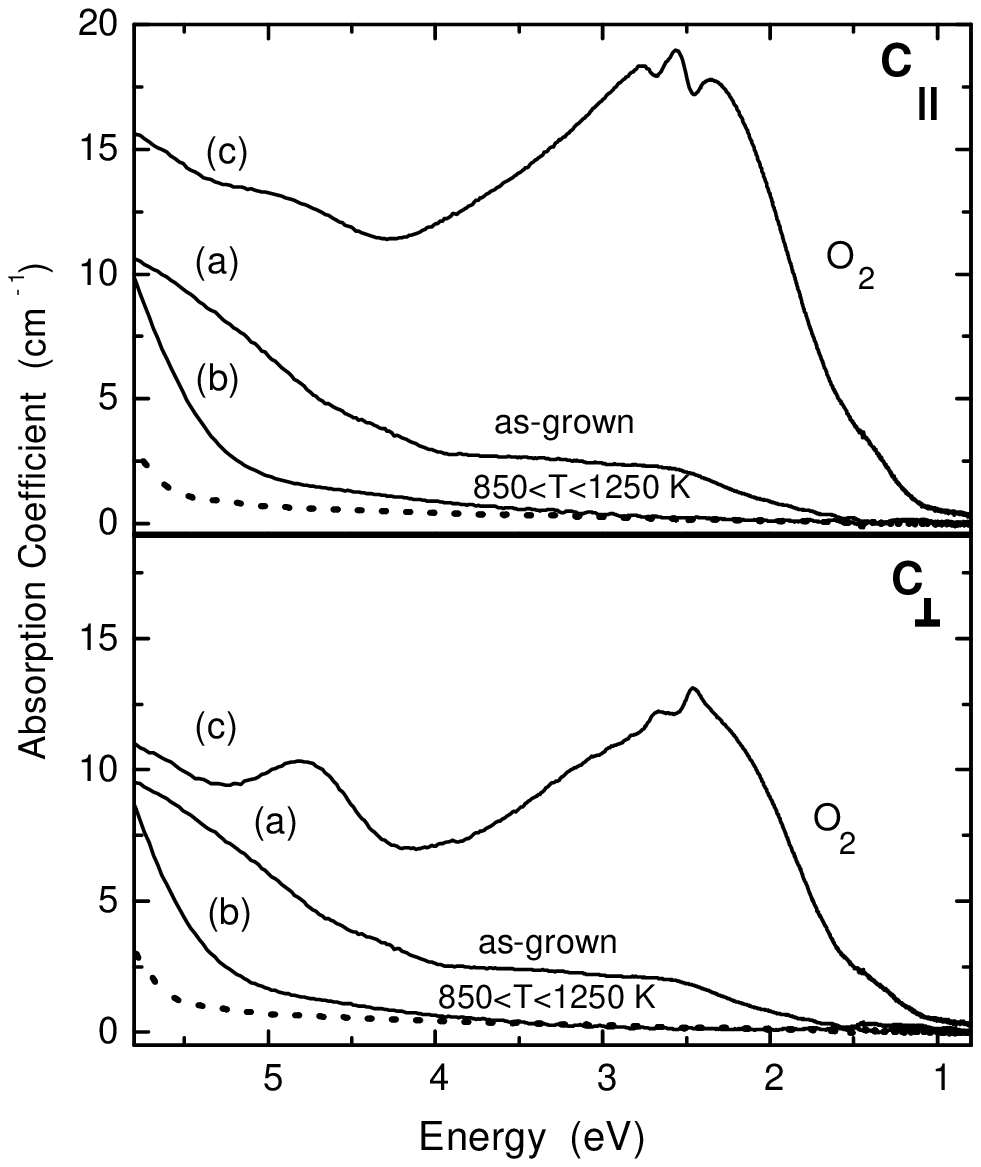}
\caption{Polarized optical spectra of an a) as-grown {\alb}
crystal, b) {\alb} crystal after annealing in air at 1200 K for
45 min, and c) {\alb} after oxidation at 1923 K for 30 min. The
dotted curve represents the spectrum of an as-grown {\alf}}
\label{fig:one}
\end{figure}
}

\newcommand{\figdos}{
\begin{figure}[th]
\centering
\includegraphics*[scale=0.7]{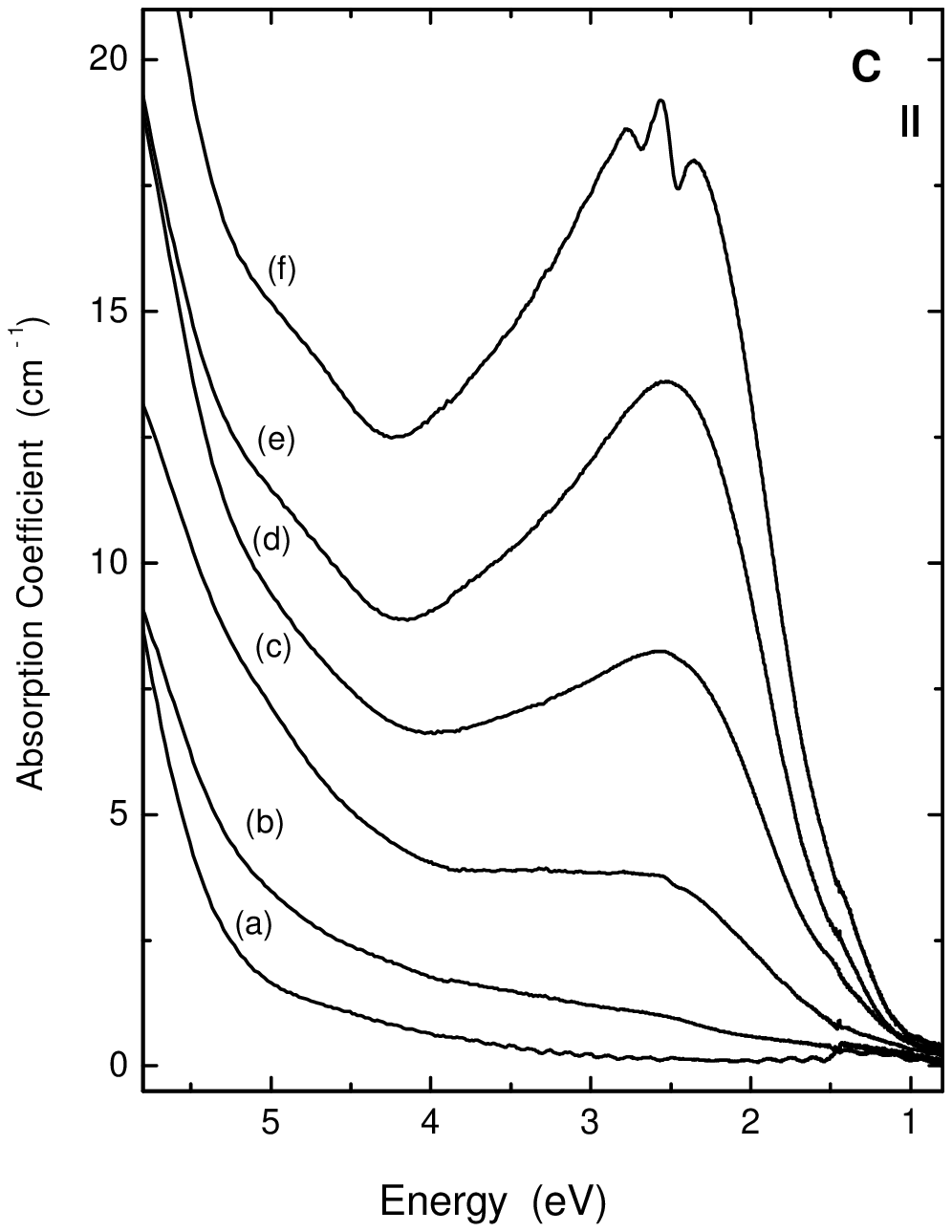}
\caption{Effect of oxidation at high temperatures. Optical
absorption spectra in {\cpa} direction of an  a) {\alb} crystal
after annealing in air at 1200 K for 45 min, and after oxidation
for 30 min  at b) 1473 K, c)1573 K, d) 1723 K, e)1823 K, and f)
1923 K.}
\label{fig:two}
\end{figure}
}

\newcommand{\figtres}{
\begin{figure}[tb]
\centering
\includegraphics*[scale=0.7]{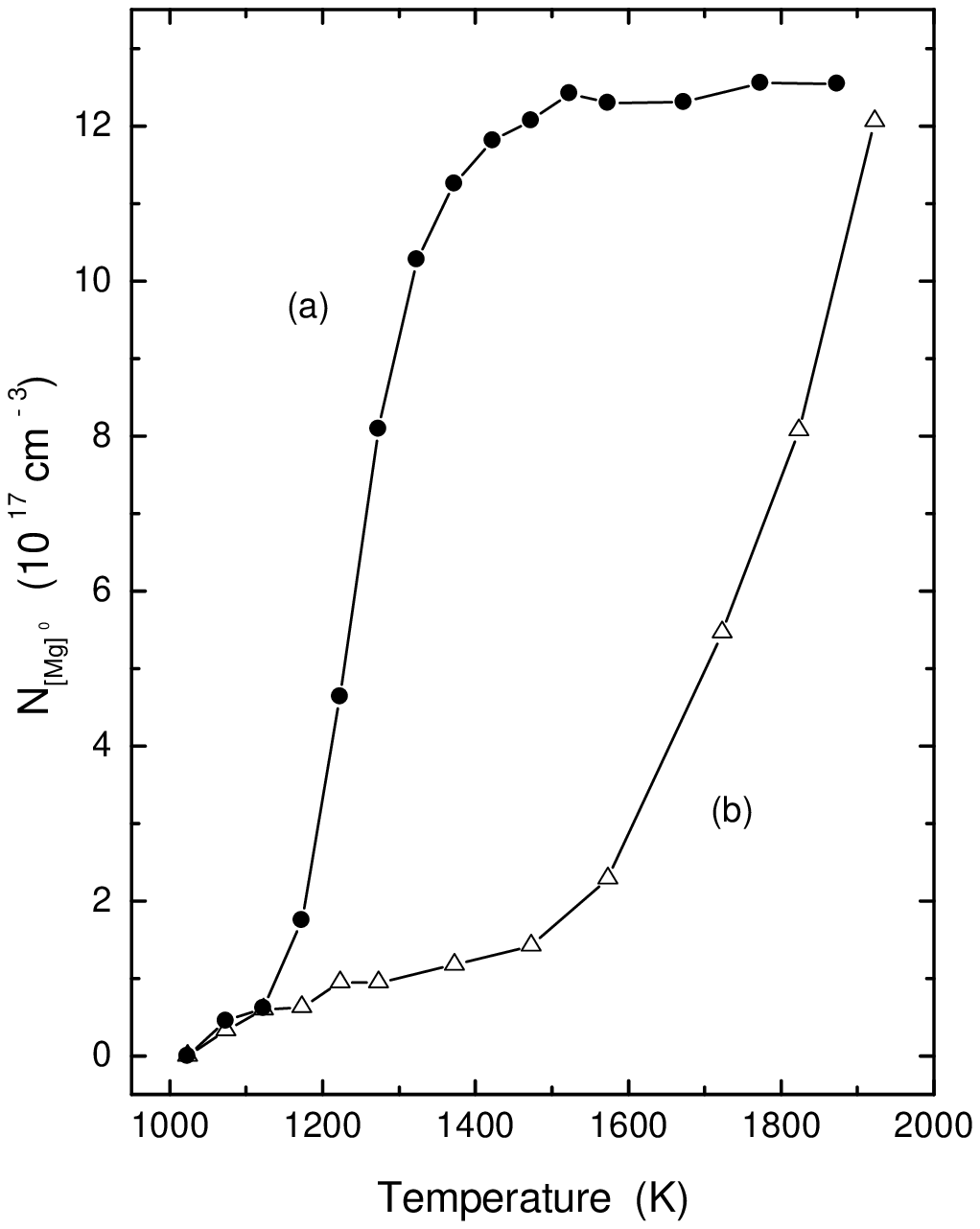}
\caption{Concentration of {\mg} centers as a function of
isochronal temperature in oxygen for (a) {\ald} sample oxidized at
1823 K for 2 h and subsequently reduced at 1223 K for 20 min, and
(b) a sample similarly reduced without pre-oxidation}
\label{fig:three}
\end{figure}
}

\newcommand{\figcuatro}{
\begin{figure}[tb]
\centering
\includegraphics*[scale=0.7]{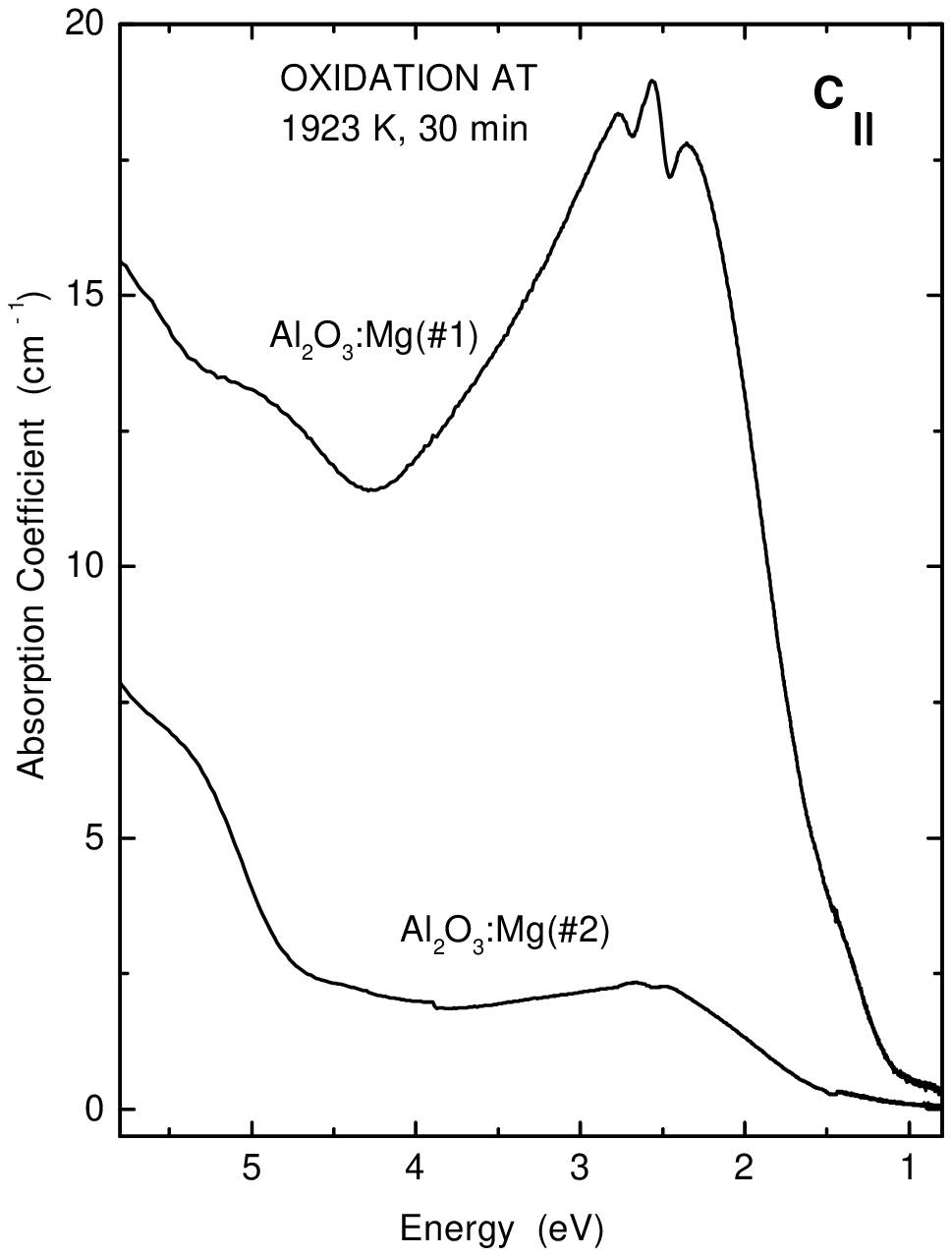}
\caption{Comparison of the optical absorption spectra of two
crystals in the {\cpa} direction: {\ald} and b) {\ale} crystal,
after oxidation for 30 min at 1923 K.}
\label{fig:four}
\end{figure}
}

\newcommand{\figcinco}{
\begin{figure}[th]
\centering
\includegraphics*[scale=0.7]{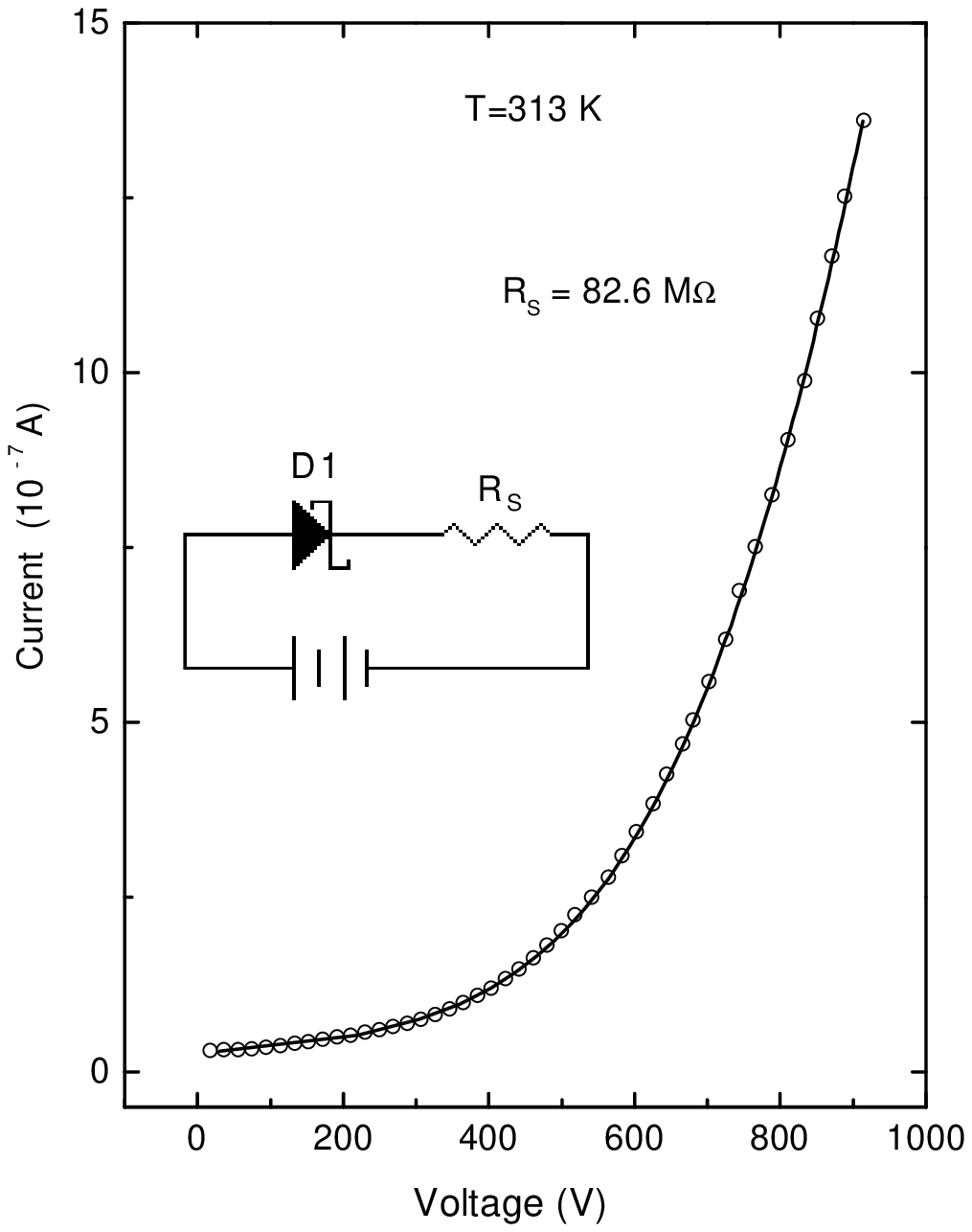}
\caption{Direct current I-V al 313 K for an {\ald} crystal
containing {\mg} centers. The solid line represents the best fit
of the experimental points to a directly biased barrier connected
to series resistance.}
\label{fig:five}
\end{figure}
}

\newcommand{\figseis}{
\begin{figure}[th]
\centering
\includegraphics*[scale=0.7]{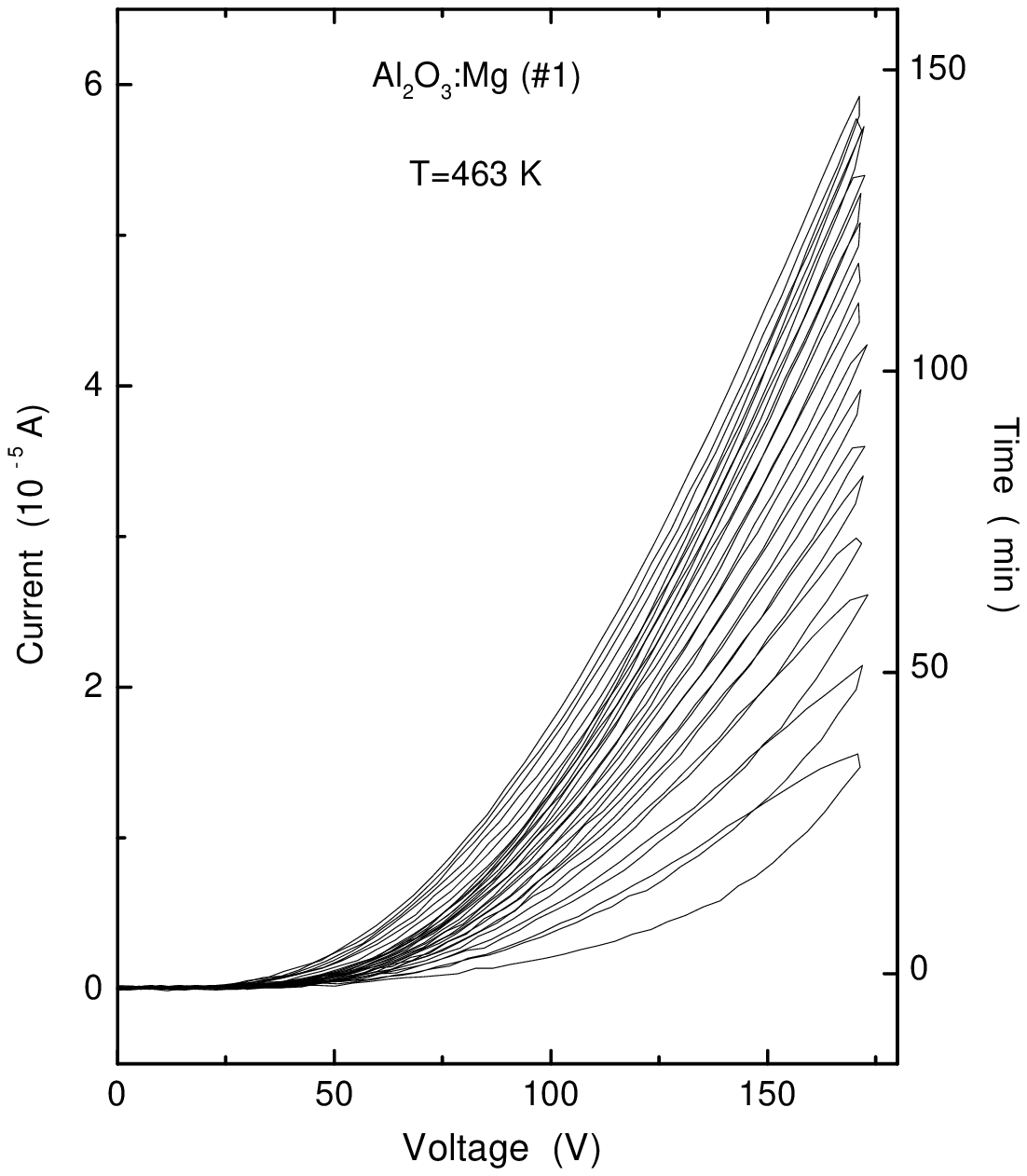}
\caption{Time evolution of the I-V characteristic at 463 K for an
{\alb} crystal containing {\mg} centers.}
\label{fig:six}
\end{figure}
}

\newcommand{\figsiete}{
\begin{figure}[th]
\centering
\includegraphics*[scale=0.7]{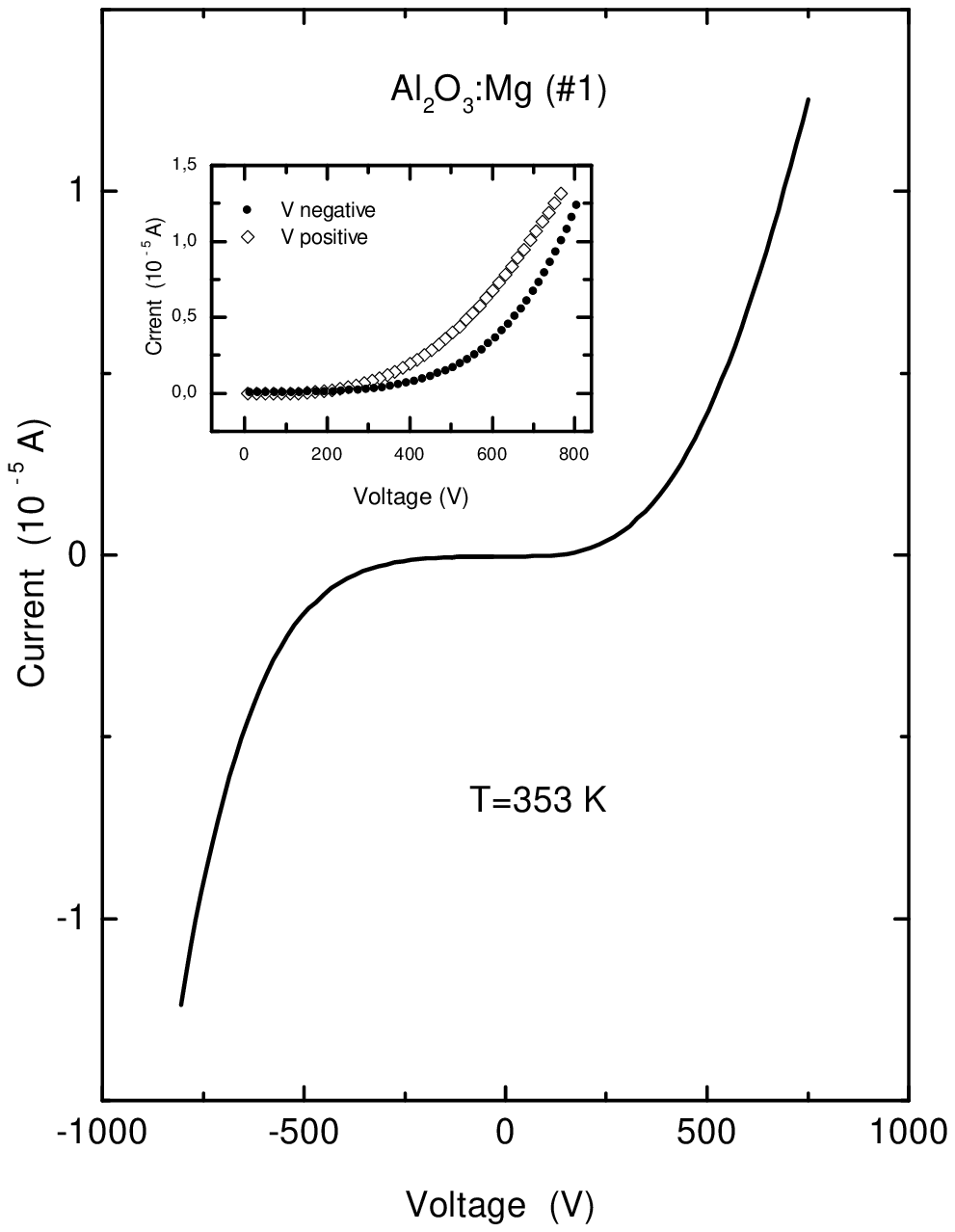}
\caption{Direct current I-V characteristic at 353 K for an {\alb}
crystal containing {\mg} centers. In the inset the positive and
negative parts of the curve are superimposed.}
\label{fig:seven}
\end{figure}}

\newcommand{\figocho}{
\begin{figure}[tb]
\centering
\includegraphics*[scale=0.7]{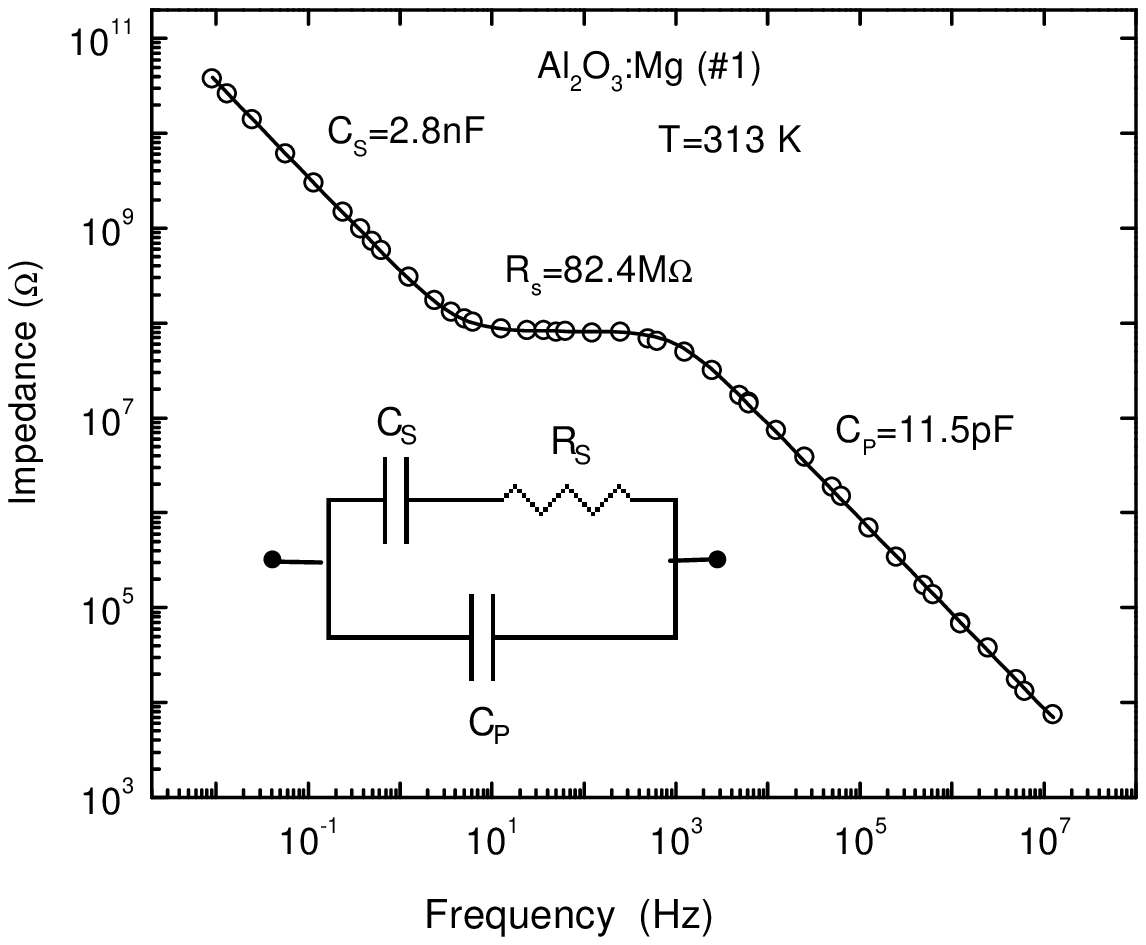}
\caption{Log-log plot of impedance against frequency at 313 K for
an {\alb} crystal containing {\mg} centers. The solid line
represents the best fit of the experimental points to the
equivalent circuit.}
\label{fig:eigh}
\end{figure}}

\newcommand{\fignueve}{
\begin{figure}[th]
\centering
\includegraphics*[scale=0.7]{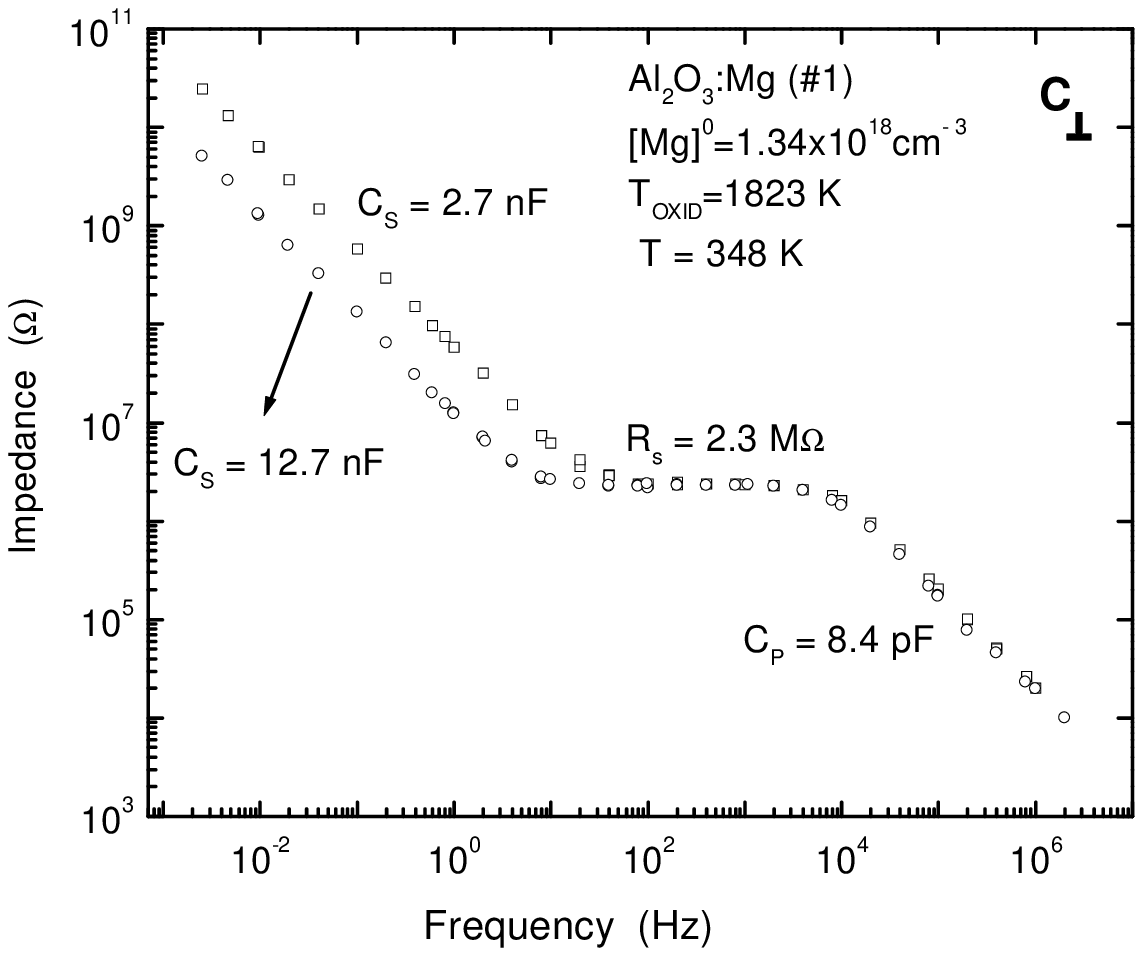}
\caption{Log-log plot of impedance against frequency at 348 K for
an {\alb} crystal containing {\mg} centers.}
\label{fig:nine}
\end{figure}}

\newcommand{\figdiez}{
\begin{figure}[tb]
\centering
\includegraphics*[scale=0.7]{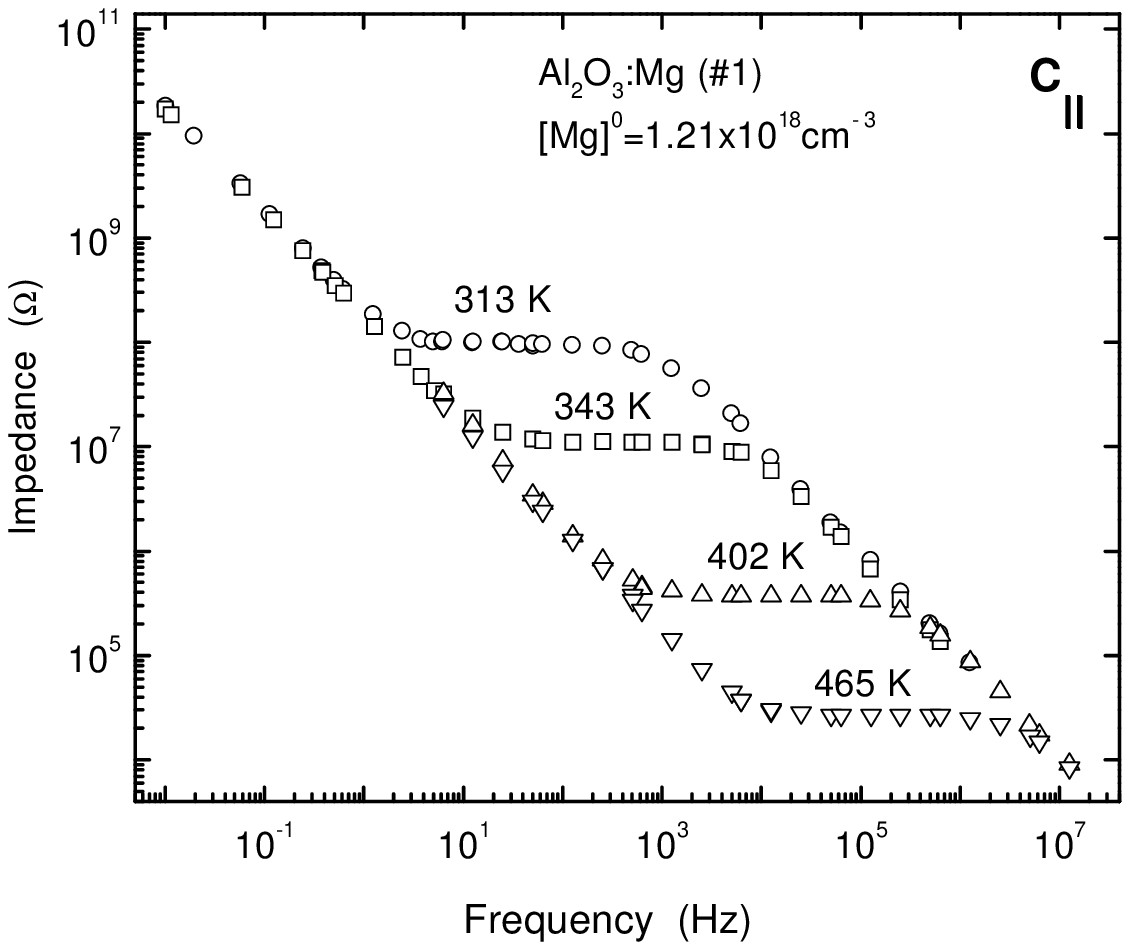}
\caption{Log-log plot of impedance {\cpa} against frequency for an
{\ald} crystal containing {\mg} centers at four different
temperatures: a) 313 K, b) 343 K, c) 402 K, and d) 465 K.}
\label{fig:ten}
\end{figure}
}

\newcommand{\figonce}{
\begin{figure}[tb]
\centering
\includegraphics*[scale=0.7]{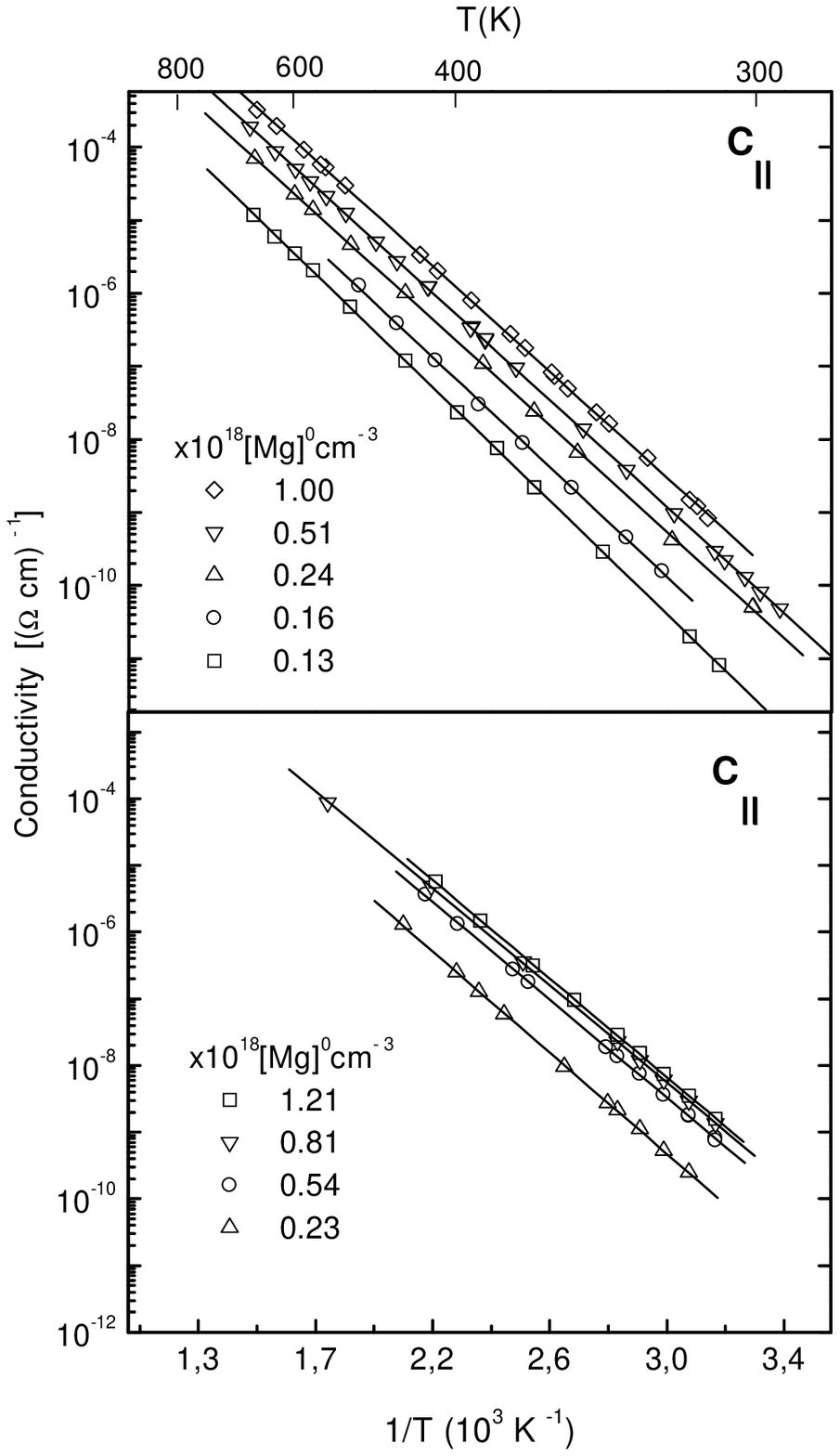}
\caption{Conductivity measured in the direction perpendicular to
the {\it c}-axis versus $T^{-1}$ for samples oxidized at the
indicated temperatures. See text for details.}
\label{fig:eleven}
\end{figure}
}

\newcommand{\figdoce}{
\begin{figure}[th]
\centering
\includegraphics*[scale=0.7]{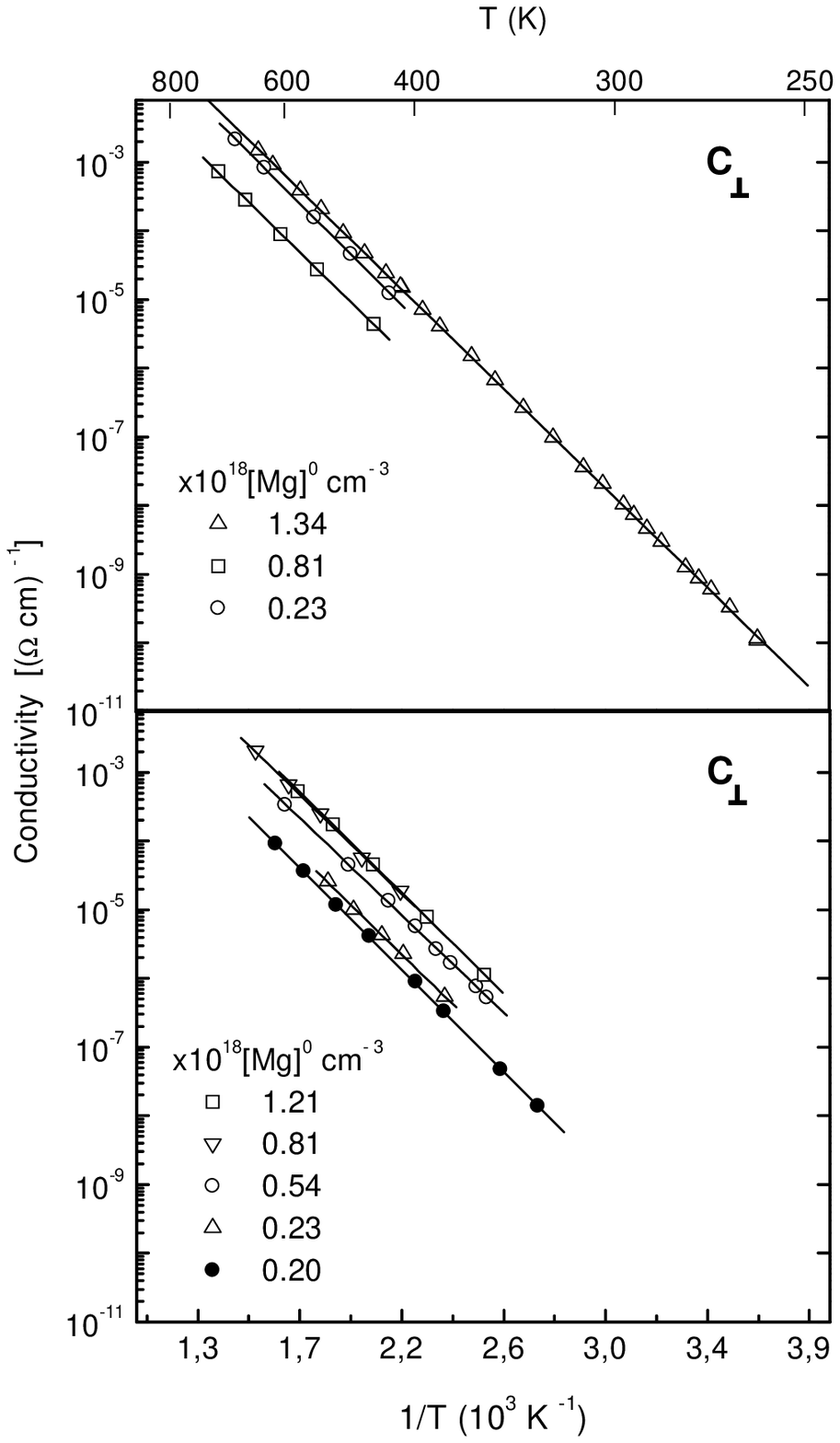}
\caption{Conductivity measured in the direction parallel to the
{\it c}-axis versus $T^{-1}$ for samples oxidized at the indicated
temperatures. See text for details.}
\label{fig:twelve}
\end{figure}
}

\newcommand{\figtrece}{
\begin{figure}[tb]
\centering
\includegraphics*[scale=0.7]{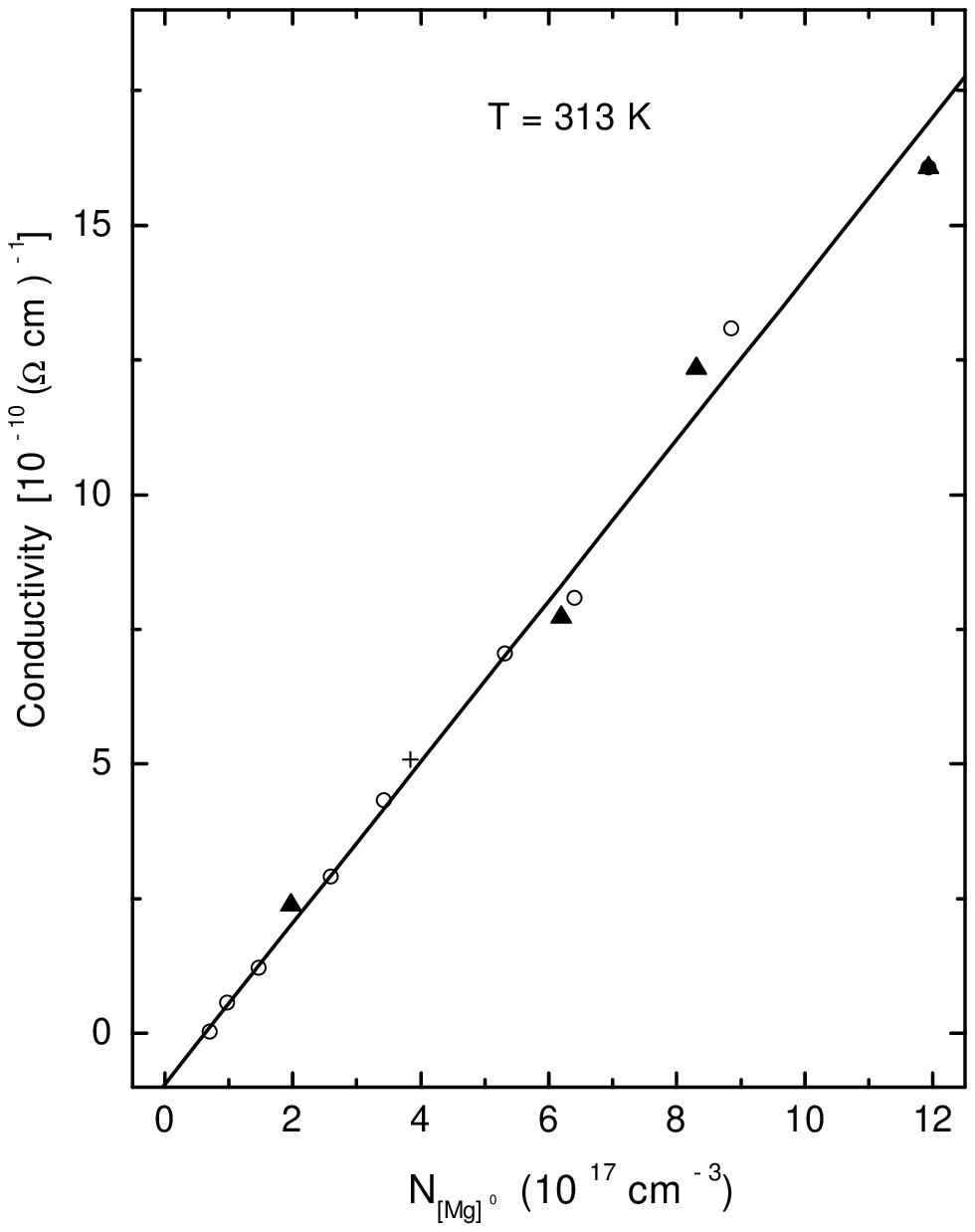}
\caption{Conductivity measured in the direction perpendicular to
the {\it c}-axis versus {\mg} centers at T = 313 K. See text for
details.}
\label{fig:thirteen}
\end{figure}
}

\newcommand{\figcatorce}{
\begin{figure}[th]
\centering
\includegraphics*[scale=0.7]{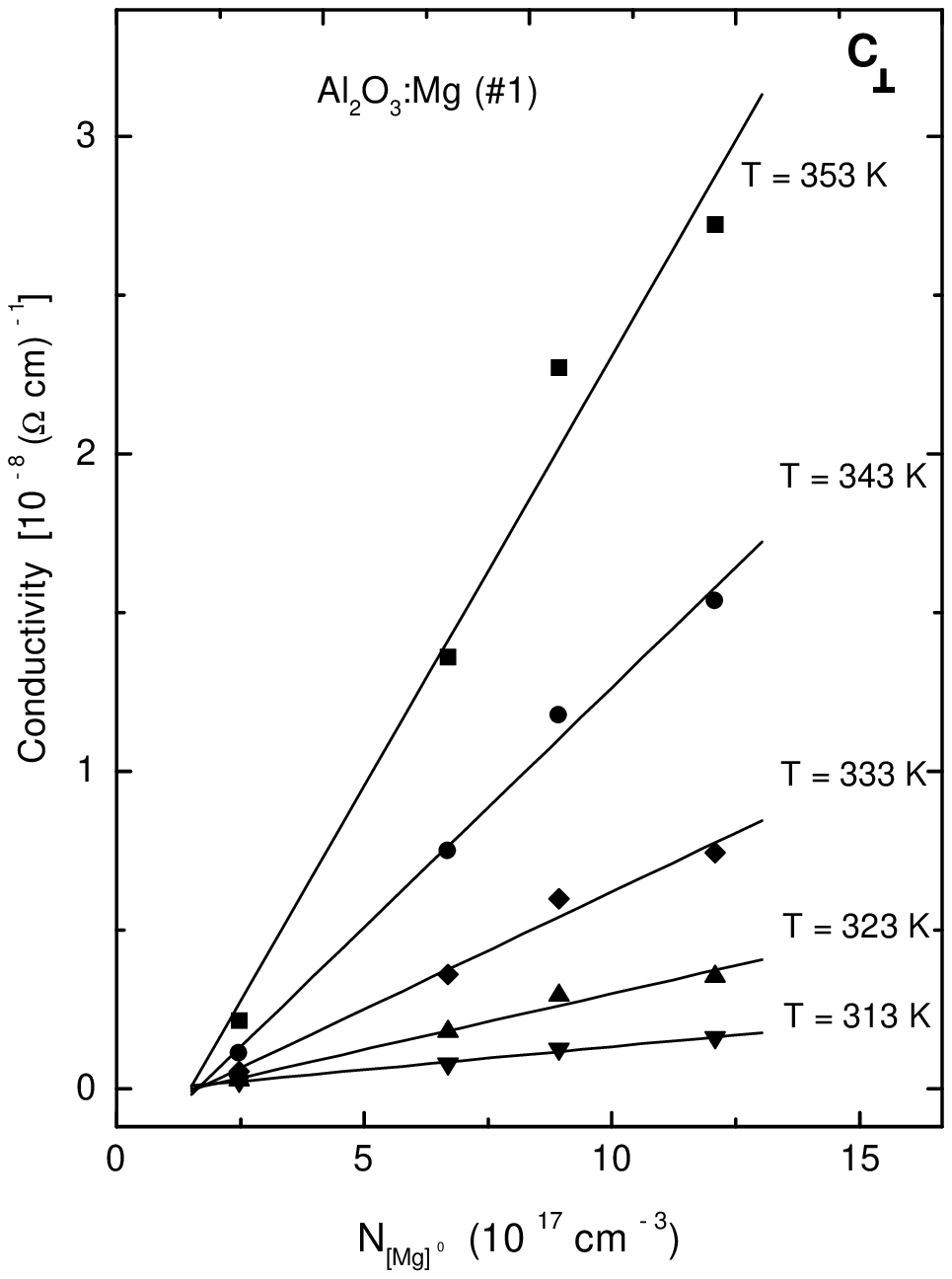}
\caption{Conductivity measured in the direction perpendicular to
the {\it c}-axis versus concentration of {\mg} centers at 313,
323, 333, 343, and 353 K.}
\label{fig:fourteen}
\end{figure}
}
\newcommand{\figquince}{
\begin{figure}[th]
\centering
\includegraphics*[scale=0.8]{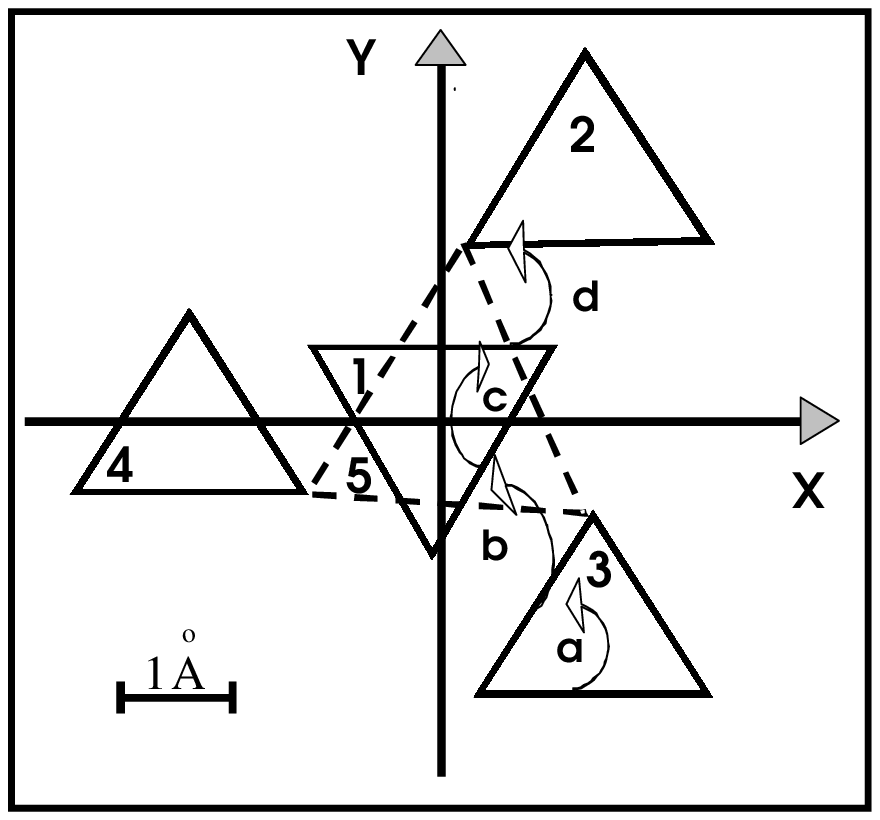}
\caption{ The {\it z} coordinates of two oxygen planes in {\alf}
crystals. Oxygen atoms are at the vertices of the {\it "small"}
triangles labeled 1,2,3 and 4. Arrows represent equal energy
hops. The path a-b-c-d involves a displacement of 4.75 \AA in the
direction perpendicular to the {\it c}-axis. A hop in the
direction parallel  to the {\it c}-axis is represented by the
arrow b (After Ref. 14).}
\label{fig:fifteen}
\end{figure}
}


\section*{I. INTRODUCTION}

{\alf} and cubic MgO crystals are well-known for their electrical
insulating properties.  In ideally pure systems one would expect
that replacing the host cation with an aliovalent impurity would
result in either electron or hole conduction. This has not been
realized, simply because of the presence of other impurities,
which serve as charge-compensators for the intended dopant.

Nevertheless, semiconducting properties have been observed in MgO
doped with lithium, and the physics involved has been reported.
\cite{uno,dos,tres,cuatro} There are similarities between {\alb}
and lithium-doped MgO crystals: 1) After oxidation at high
temperatures, hole-trapped centers are formed in both oxides. In
the latter, {\lic} centers (substitutional {\li} ions, each
attended by a hole) are produced at T $>$1100 K, which absorb
light at 1.8 eV (690 nm).\cite{cinco,seis,siete} 2) In MgO:Li
crystals, hole-trapped centers are also responsible for the
electrical conductivity enhancement observed.~\cite{cuatro} MgO
crystals containing {\lic} centers have been shown to be p-type
semiconducting at temperatures not much higher than RT, with an
acceptor level of 0.7 eV.~\cite{cuatro} By analogy, {\ala}
crystals are expected to serve as a p-type semiconductor. Both
systems have potential as high-temperature semiconductors.
However, whereas the former is brittle and will have limited
applications, the latter is expected to have superior mechanical
integrity.

As-grown single crystals of magnesium-doped {\alf} exhibit a
gray-purple coloration, which becomes much stronger after
quenching these crystals from above 1173 K in an oxidizing
atmosphere.~\cite{ocho,nueve,diez} This coloration is due to a
broad asymmetric optical absorption band, centered at $\approx$
2.5 eV (496 nm), and has been attributed to the paramagnetic
hole-trapped {\mg} center, which is a substitutional {\mgi} ion
with a trapped hole localized on one of the six NN oxygen
ions.~\cite{ocho,nueve} These centers are stable up to 850 K. In a
reducing atmosphere {\mg} centers vanish at temperatures above
$\approx $ 850 K. Thermal treatments between 850 K and 1250 K in
air turn the samples colorless.

Nominally pure {\alf} has a reported~\cite{once,doce}
conductivity varying from $10^{-18}$ {\ohc} at room temperature to
$10^{-9}$ {\ohc} at 1773 K. The effect of magnesium doping on its
electrical conductivity has been previously ~\cite{nueve,trece}
studied at elevated temperatures (1673-1873 K ), and the value
reported at 1773 K is $10^{-5}$ {\ohc}; this increase in
conductivity was attributed to holes released from {\mg} centers
moving as {\it large} polarons with an activation energy of 0.68
eV.~\cite{nueve} However, recent theoretical
calculations~\cite{catorce} in {\alf} suggest that holes exist as
small polarons and their optical absorption energy is $\approx$
2.9 eV. The energies required for polaron hopping between
neighboring positions is $\approx$ 0.9 eV.~\cite{catorce}

In the present work, the electrical properties of {\alb}
containing {\mg} centers were studied at moderate temperatures
(from 250 to 773 K), i.e. at temperatures where {\mg} centers are
stable. Both AC and DC electrical measurements were performed. At
RT, the resulting values for the conductivity are ten orders of
magnitude higher than those reported for undoped samples. The
conductivity dependence on {\mg} content was investigated. The
effect of the temperature on the conductivity along different
crystallographic directions was analyzed, and provided
information about the conductivity mechanisms. Our results favor
the {\it small} polaron model.~\cite{catorce}

\section*{II. EXPERIMENTAL PROCEDURE}

The {\alc} single crystals were grown separately by the
Czochralski method (the ambient growth atmosphere was an inert
gas with $\approx$ 1000 ppm by volume of oxygen), labeled as
{\ald} and {\ale}. The latter had a higher magnesium
concentration, but the concentration of other impurities was also
higher. Measurements were primarily made in the {\ald} crystal.
{\ale} crystal was used for comparison. Atomic Emission
Spectrometry analyses indicated that for {\ald} crystal the
magnesium concentration was 25 ppm; other known impurities were
Ti ($<$ 10 ppm), V ($<$ 3ppm), Cr ($<$ 3 ppm), Mn ($<$ 5 ppm), Fe
(10 ppm), and Ni ($<$ 5 ppm). For crystal {\ale}, the magnesium
content was 57 ppm and other impurity concentrations were: Ti (35
ppm), V ($<$ 3ppm), Cr (10 ppm), Mn (5 ppm), Fe (60 ppm), and Ni
(6 ppm). Samples of about 1 mm thick were cut from the boule with
either the {\it c}-axis parallel or perpendicular to the broad
face, and polished to optical transparency. Optical absorption
measurements were made with a Perkin-Elmer Lambda 19
Spectrophotometer.

Heat treatments either in static air or in flowing oxygen were
made with the samples placed in a platinum basket inside an
alumina tube inserted in a horizontal furnace. Reduction
treatments were performed in a high-purity graphite capsule
surrounded by flowing nitrogen gas inside the furnace.
Subsequently, the samples were pulled out from the hot zone and
fast-cooled in the room-temperature portion of the tube.

For DC measurements, voltage was applied to the crystals with a
DC Sorensen DCS 150-7 voltage source. I-V characteristics were
measured with an electrometer (Keithley 6512) and a voltmeter (HP
34401A). A standard three electrical-terminal guard technique was
used.~\cite{quince} For AC measurements, a function generator was
used, where available frequencies range from $10^{-4}$ to
$10^{7}$ Hz.

Electrodes were made by sputtering metals with different work
functions (Al, Mg and Pt) onto the  sample surfaces, usually the
largest faces. The electrical response was independent of contact
electrode materials, so mostly Al electrodes were used. The
samples with the electrodes were placed in a sample holder, which
can be used for measurements below room temperature (using a
temperature controlled thermal bath) and for temperatures above
room temperature using a conventional horizontal furnace.
Electrical measurements were performed in static air. The
temperature of the sample was monitored with a Chromel-Alumel
thermocouple in direct contact with the sample.

\section*{III. EXPERIMENTAL RESULTS}
\subsection*{A. Characterization by optical absorption}

The polarized optical absorption spectra of a sample from {\ald}
before and after different thermal treatments are shown in Fig.1.
In the as-grown state (trace a), a broad asymmetric absorption
band covering the visible region is observed. The high energy
side of this band extends into the UV part of the spectrum. The
low energy side, extending into the infrared, has an
approximately Gaussian shape peaking at 2.56 eV. This band is the
analog of the {\lic}  center in MgO:Li and has previously been
associated with {\mg} centers.~\cite{ocho,nueve} The presence of
these hole-trapped defects in as-grown samples is indicative of
the presence of oxygen during the crystal growth process. Indeed,
annealing of the latter crystals in a reducing atmosphere (or in
air) at 1200 K for 45 min, completely annihilates this band
(trace b). For comparison, the absorption spectra of an as-grown
undoped {\alf} crystal are also shown in the dotted curves. The
higher absorption in the UV region  is probably due to
charge-transfer bands of transition metal ions.~\cite{dieciseis}

\figuno

After oxidation of the as-grown crystal at 1923 K for 30 min,
the  absorption of the 2.56 eV dramatically increased and another
band at about 4.75 eV (260 nm) emerged, which can be better
resolved in the spectrum taken with light polarized perpendicular
to the  {\it c}-axis. The structure on top of the 2.56 eV band is
better resolved as the oxidation temperature is increased. Its
origin is unknown. In addition, the intensity of the 2.56 eV band
is higher for the spectrum with light propagating parallel to the
 {\it c}-axis. The ratio between the two absorption coefficients is
$\approx $ 1.5.

The concentration of {\mg} centers can  be  estimated from the
experimentally deduced~\cite{nueve} relationship

\begin{equation}
\label{eq:absorption}
\hspace{0.5cm}N(cm^{-3})\approx1.14\times10^{17}\alpha_{\bot}(cm^{-1})
\end{equation}

where N({\cm}) denotes the concentration of {\mg} centers and
$\alpha( cm^{-1}$) is the absorption coefficient of the 2.56 eV
band. The resulting concentration for the sample oxidized at 1923
K is 1.20 $10^{18}$ {\cm} or 12 ppm, indicating that
approximately half of the magnesium is dissolved in the crystal
in the {\mg} configuration.

Another sample from {\ald}, initially annealed in air to remove
the 2.56 eV band, was oxidized for 30 min periods at increasing
temperatures in flowing oxygen up to 1923 K. Fig.2 shows the
growth of the {\mg} center band for several selected
temperatures. After the anneal at about 1100 K, the 2.56 eV band
began to emerge. The absorption increases monotonically with
temperature up to about 1923 K and then saturates. The UV
absorption at 4.75 eV, which is more pronounced in the {\it c}
direction, also increases with the oxidation temperature. We
associate this band with a Mg-related defect because undoped
crystals or those doped with other impurities do not show this
band  after oxidation at elevated temperatures.

\figdos

Prior thermal history either in reducing or oxidizing atmosphere
has a strong influence on the formation rate of {\mg} centers
induced by oxidation. Two similar samples from the {\ald} crystal
were subjected to thermal treatments in different atmospheres. One
as-grown sample was reduced in graphite at 1223 K for 20 min
until the {\mg} centers vanished. The second as-grown sample was
first oxidized at 1723 K for 2 h before reducing at 1223 K for 20
min. These two samples were then isochronally heated in flowing
oxygen for 30 min (see Fig.3). The threshold temperatures for the
formation of {\mg} centers are the same in both cases $\approx$
1050 K. However, the production curve is much more rapid for the
pre-oxidized sample (curve a). The concentration saturates at
$\approx $ 1500 K with $ \approx$ 1.3 $10^{18}$ {\cm}. Without
pre-oxidization (curve b) the same concentration was attained
only after 1923 K, which is the maximum capability of our furnace.

\figtres

A sample from the {\ale} crystal, which has twice the
concentration of magnesium as from {\ald}, generates fewer {\mg}
centers upon oxidation (Fig.4). Whereas in sample ${\#1}$, 50{\%}
of the magnesium forms {\mg} centers, in sample ${\#2}$ it was
only 3{\%}. This finding clearly shows that the resulting {\mg}
concentration depends on the content of magnesium as well as
other impurities. Presumably, other impurities, mainly Ti and Cr,
prevent the trapping of holes by oxygen ions located as nearest
neighbors of the {\mgi} ions. Higher oxidation states such as
{\Ti} and {\Cr} compensate {\mgi} ions substituting for {\ali}
ions. Indeed, the {\ale} sample has a relatively higher
absorption in the UV region.

\figcuatro

\subsection*{B. Electrical Measurements}

To investigate the electrical conductivity of {\alb} crystals, DC
and AC electrical measurements were performed in samples with
different concentrations of {\mg} centers in the temperature
range 250-800 K. Unless otherwise indicated, electrical
measurements were made in {\ald} samples where oxidation induces
a higher concentration of {\mg} centers.

\subsubsection*{1. DC electrical properties}

DC electrical measurements at low electric fields reveal blocking
contacts. However, at high fields, the reverse bias
characteristic is that of a "soft" leaky barrier. A typical
forward-bias current-voltage (I-V) characteristic is shown in
Fig.5 for a sample with a {\mg} concentration of
$\approx$1.20$\times10^{18}$ {\cm}. This characteristic is similar
to that of a diode with a series resistance, {\Rs}, which
corresponds to a blocking contact at one side of the sample and
an ohmic contact at the other side; the series resistance is the
bulk resistance of the sample. The experimental points are
plotted as open circles, and the solid line represents the best
fit of the data to the equation:

\begin{equation}
\label{eq:IV}
\hspace{0.5cm}V=\frac{nkT}{q}\times\ln(\frac{I}{I_{s}}+I\times R_{s})
\end{equation}

which corresponds to a forward biased diode in series with a
resistance. Here {\it n} is the ideality factor of the junction,
{\it q} is the carrier charge, and {\it $I_{s}$} is the saturation
current.

\figcinco

Blocking contacts are expected for hole conduction in wide band
gap insulators, but since the electrodes are symmetric (the metal
is the same in both), there should be no ohmic contact. On
reversing the applied voltage, practically the same I-V
characteristic is obtained. This result appears to be  in
contradiction with having an ohmic contact on one side and a
blocking contact on the other: a given polarity would imply a
forward bias in the barrier (a direct current flows through the
sample), whereas the opposite polarity should induce a reverse
bias in the barrier, and small and voltage-independent values for
the current should be observed.

Symmetry requires that blocking contacts are formed on both sides
of the sample. In this way, regardless of the polarity of the
applied voltage, one of the contacts is forwardly biased, while
the other is reversely biased; on increasing the applied voltage,
breakdown can occur in the reversely biased contact. For
sufficiently high voltages, the overall behavior of the sample is
that of an ohmic contact with a series resistance (the bulk
resistance) and a forward biased blocking contact. Three
parameters can be obtained from a fit of the experimental I-V
characteristics: the bulk resistance, the ideality factor of the
junction, and the saturation current. The values of the last two
parameters are affected mainly by the shape of the low voltage
part of the I-V curve. On the other hand, the initial part of the
characteristic is probably influenced by the breakdown
peculiarities in the reversed direction and the values derived
for {\it n} and {\it $I_{s}$} are not fully reliable. Regarding
the value obtained for the series resistance, it provides an
order of magnitude for the sample, as the AC results following
this section will show. The association of the ohmic part of the
I-V curve with the sample resistance is based on the fact that the
same value for the conductivity $\sigma= d/SR_{s}$ was determined
using samples with different thickness, {\it d}, and
cross-section, {\it S}.

As the temperature is raised, the bulk resistance diminishes; it
is not feasible to apply the high voltages needed to break the
blocking contact because of Joule dissipation in the sample. At
medium voltages, the rupture mechanisms of the blocking contact
reveal a complex behavior; since neither the electrodes nor the
sample surface is homogeneous on a small scale, some regions of
the contact can be broken at medium voltage values, while others
depend on temperature and time the voltage is applied. This
behavior is illustrated in Fig.6 for a sample subjected to
subsequent voltage ramps (the right axis represents the time at
the maximum applied voltage). The slope of the I-V curve
increases with time, implying that the resistance diminishes;
also there is some hysteresis due to Maxwell-Wagner polarization
phenomena at shorter times.~\cite{diez,diecisiete,dieciocho} These
two observations indicate that the area of the broken contact
increases with time, thereby diminishing the total resistance and
the polarization effects. If the whole contact is broken, the
resistance should be equal to the sample resistance.

\figseis

\figsiete

Additional information on the non-homogeneity of the contacts is
shown in Fig.7. The I-V curve is non-symmetric with respect to
voltage polarity, probably because the size of the contact
regions that experience breakdown is different on each electrode.
To obtain symmetric curves special care must be taken to assure
that sample preparation and electrode deposition are identical in
both electrodes. Another important outcome of the DC measurements
is the emission of electroluminescence (EL): when a current in
excess of $\approx$10 {\mA} passes through the sample, steady
light has been observed with the naked eye to emerge from a narrow
region at the negative electrode~\cite{diecinueve}. No emission
was observed from the positive electrode. The light intensity is
proportional to current intensity above the threshold. Several EL
bands have been resolved and related to photoluminescence
emissions of different extrinsic defects, such as {\Cr} ions.
These results have been interpreted based on the assumption that
the majority carriers flowing through the sample recombine with
the minority carriers injected from the negative
electrode~\cite{diecinueve}. EL emission occurring at the
negative electrode indicates that the majority carriers are
holes. The steady emission and its proportionality with current
and voltage reveals that ohmic conduction is taking place.

As mentioned in the experimental section, we used three metals as
electrodes, Mg, Al and Pt, with widely different work functions:
3.6, 4.2 and 5.6 eV, respectively. The results described in this
section are independent of the types of metal used as electrodes.
This indicates that the barrier formed at the electrodes is
dominated by surface effects.

The relevant conclusion of DC measurements in {\alb} crystals
containing {\mg} centers is that upon application of a moderate
electric field there is a flow of direct current through the
sample, which is ohmic in the high voltage regime, and is
governed by the bulk resistance. Surface and electrode effects
make the interpretation of these measurements difficult and
sometimes not quantitatively reproducible. Some of these
difficulties can be overcome by AC electrical measurements.

\subsubsection*{2. AC electrical properties}

The equivalent circuit for low voltage AC measurements consists
of {\Rs} in series with {\Cs} (junction capacitance, which
accounts for the blocking nature of the contacts) and {\Cp}, which
represents the dielectric constant of the sample (see Fig.8
insert).

\figocho

Fig.8 is a log-log plot of the AC impedance versus frequency for
a sample with an electrode area of  0.85 {\cma} and a thickness
of 0.10 cm, at a fixed temperature of 313 K.  The experimental
values are plotted as open circles, and the solid line is the
best fit to the equivalent circuit. A value of 11.5 pF for {\Cp}
is consistent with the dielectric constant of {\alf}. The results
presented in Figs.4 and 7 were measured in the same sample and at
the same temperature; the values determined for {\Rs} in both DC
and AC experiments are in good agreement.

The values obtained for {\Cs} depend on the quality of the sample
surface. Fig.9 shows two log-log plots of impedance versus
frequency in a sample with a concentration of {\mg}centers of $
\approx$ 1.34 $\times 10^{18}$ {\cm} ({\it S} = 0.75 {\cma} and
{\it d} = 0.12 cm). The two curves correspond to the sample with
two degrees of surface polishing; the values of {\Cs} = 2.7 and
12.7 nF were obtained for the sample with an intermediate and
final polishing, respectively. After a DC current flowed through
the sample for 24 h at a constant temperature of 348 K, the {\Cs}
value dropped to 7.2 nF. Subsequently, the sample was heated at
600 K for several hours with a small AC applied voltage and {\Cs}
dropped further to $\approx$ 2.7 nF. Since {\Cp}  is more than
one thousand times smaller (8.4 pF), a value of 12.7 nF for {\Cs}
suggests a depletion layer thickness slightly smaller than one
micron, which seems to be reasonable. The diminishing {\Cs} with
the flow of DC current is attributed to the presence of traps in
the depletion layer. The hysteris behavior of {\Cs} ({\Cs} does
not recover its original 12.7 nF value) can be attributed to deep
traps at the depletion layer. Unfortunately, characterization of
these traps, for instance by deep level transient spectroscopy
(DLTS), is prohibitive since the high impedance of the sample
requires the use of very low frequencies for these measurements.
In addition, the ionization temperature of these traps would
require that the sample be heated at temperatures at which the
{\mg} centers are unstable. Lastly, the {\Cs} values are observed
to be independent of the metal used as electrodes.\\

\fignueve

Fig.9 also shows the frequency regions in which the resistance
and parallel capacitance, {\Rs} and {\Cp} dominate ($>$ 102 Hz).
These values are independent of the quality of the sample
surface. The {\Rs} values in the AC measurements are reproducible
and yield consistent values for the sample conductivities.

We have also tried to perform a {\it {\Cs}(V)} characterization,
the dependence  of the  junction capacitance on the voltage bias.
Within the voltage range (0-100 V) investigated, {\Cs} is
independent of the voltage bias. However, these measurements are
difficult because low frequency values are necessary. On the
other hand, and of greater importance, the barrier is "leaky", so
upon application of a bias voltage some regions of the barrier
junction are broken. Thus, the results of low-voltage AC
measurements reinforce the interpretation from DC measurements
that blocking contacts exist in a resistive sample.

The temperature dependence of the three parameters {\Cs}, {\Cp}
and {\Rs} can be deduced from Fig.10. AC measurements at four
different temperatures were made in a {\alb} sample with a
concentration of $\approx$ 1.21$ \times 10^{18}$ {\cm} {\mg}
centers. The basic results are: {\Cs} and {\Cp} are practically
independent of temperature in the temperature range 273-423 K,
whereas the sample resistance diminishes as temperature increases.

\figdiez

Next, we will address the electrical conductivity and its
dependence on temperature and {\mg} concentration. We have
already mentioned that experiments using different sample
geometries confirm that the resistance {\Rs} is directly related
to the sample conductivity, thus the dependence of the
conductivity on temperature and {\mg} content can be inferred
from {\Rs} measurements. Figs.11 and 12 show respectively the
Arrhenius plots of the sample conductivity {\it perpendicular} and
{\it parallel} to the crystallographic {\it c}-axis for samples
containing different concentrations of {\mg} centers. The data
plotted in the top part of these figures were obtained from
different samples cut from the {\ald} crystal, while the bottom
part refers to only one sample (also from {\ald} subjected to
oxidizing treatments at increasing temperatures. After each
anneal the AC measurements were performed at different
temperatures. The solid circles in Fig.12 are data for an {\ale}
sample oxidized at 1923 K for 30 min. The slope of these plots is
the same, indicating that the conductivity is thermally activated
with an activation energy of $\approx$ 0.68 eV.  The parallelism
of the straight lines indicates that this value is independent
of: 1) the {\mg} content, 2) the crystallographic orientation,
and 3) the amount of other impurities. However, the conductivity
depends on the crystallographic direction, and is four times
larger in the direction parallel than in the directions
perpendicular to the {\it c}-axis.

\figonce

\figdoce

Fig.13 shows the dependence of the conductivity on {\mg} center
concentration at T = 313 K. The conductivity was measured
perpendicular to the {\it c} axis. The four solid triangles
correspond to one sample from {\ald} oxidized at four isochronal
temperatures for 30 min each. Open circles represent the data
obtained from eight samples also cut from the {\ald} crystal.
Each sample was oxidized at a different temperature to produce a
different concentration of {\mg} centers. Lastly, the cross
corresponds to a sample cut from the  {\ale} crystal and
subsequently oxidized at 1923 K for 30 min. These results clearly
show that there is a linear relationship between conductivity
and  {\mg} content,  regardless of  the thermal history  and
content of other impurities.

\figtrece

To explore this linearity further, the electrical conductivity
dependence on the {\mg} content was investigated in the same
sample at five temperatures: 313, 323, 333 343 and 353 K. The
conductivity perpendicular to the {\it c} axis was measured. The
{\mg} concentration was varied by subjecting the sample to
increasing temperatures in flowing oxygen. After each oxidation,
the amount of {\mg} centers in the sample was measured, and its
conductivity was determined at the five temperatures indicated.
The results are plotted in Fig. 14. In all the cases, a linear
relationship between conductivity and {\mg} concentration was
also observed.

\section*{IV. DISCUSSION}

Our experimental results show that the electrical conductivity of
{\alc} crystals is governed by {\mg} centers in direct proportion
with the concentration. The presence of other impurities serves
to reduce the resulting {\mg} concentration. The conductivity is
thermally activated with an activation energy of 0.68 eV. The
conductivity anisotropy factor is four with {\cpa} greater than
{\cpe}.

The thermally-activated behavior of conductivity can be explained
by any of three different mechanisms: impurity
conduction,~\cite{vente,ventiuno} standard semiconducting
behavior~\cite{ventidos} or small polaron
motion.~\cite{ventitres,venticuatro} We shall discuss these three
mechanisms in relation to the electrical conductivity results.

{\it Impurity conduction} may be ruled out because this mechanism
implies a strong dependence of the activation energy on impurity
content (in this case {\mg} centers) and on the concentration of
compensating impurities.~\cite{vente,ventiuno} Our results give
the same activation energy over a broad range of {\mg}
concentration. In addition, samples with large Mg concentration
compensated by large concentrations of compensating impurities
{\ale} yield the same activation energy.

The standard {\it semiconducting mechanism}
yields~\cite{ventidos} a thermally activated behavior for the
concentration of carriers, {\it p}, in two different regimes:

a)  At low temperatures, {\it p $<< N_{D}$}  and {\it p $<<
N_{A}-N_{D}$}; here {\it $N_{A}$} and {\it $N_{D}$} are the
concentrations of acceptor and compensating impurities,
respectively, and $N_{V}$ the effective density of states in the
valence band. The concentration of carriers is given by:

\begin{equation}
\label{eq:concentration}
\hspace{2.5cm}p(T)=\frac{N_{V}(N_{A}-N_{D})}{2N_{D}}\cdot exp
{(-\phi/ kT)}
\end{equation}

where $\phi$ is the acceptor ionization energy.

b)  When the temperature is high enough for {\it p $>> N_{D}$}
and {\it p $<< N_{A}-N_{D}$}, but low enough for not having all
the acceptors ionized (kT $<< \phi$), then:

\begin{equation}
\label{eq:concentration1}
\hspace{3.5cm}p(T)=\sqrt{\frac{N_{V}(N_{A}-N_{D})}{2}}\cdot
exp{(-\phi/ 2kT)}
\end{equation}

The conductivity in both cases is  $\sigma =p \mu e$, where $\mu$
is the mobility, which usually varies as a power of the
temperature (depending on the carrier scattering process) and is
independent of the acceptor concentration for concentrations that
are not too large. In the second temperature regime, the
conductivity is proportional to the square root of the acceptor
concentration at a given temperature, which is not our case. So
we are left with the behavior given by Eq. 3, which provides a
linear relation between conductivity and acceptor concentration.
This result agrees with our findings. However, according to Eq.
3, the slope of the conductivity plotted against acceptor
concentration depends on the concentration of compensating
impurities. We know that these concentrations are very different
in {\ald} and {\ale} crystals. Fig. 14 shows that the slope of
the conductivity versus $T^{-1}$ is independent of the
concentration of compensating impurities.

\figcatorce

Within this framework it is also difficult to establish a link
between the activation energy of the conductivity and the optical
absorption energy associated with {\mg} centers (0.68 eV versus
2.56 eV). It should be noted that for ionic crystals, polar
optical scattering of carriers may be the dominant process in the
mobility, leading to a temperature dependence of the mobility
given by:

\begin{equation}
\label{eq:mobility} \hspace{3.5cm}\mu(T)=\mu_{0}\cdot
exp{(\frac{\hbar\omega_{o}}{kT})}
\end{equation}

where $\omega_{0}$ is the angular frequency of an optical phonon
and $\hbar\omega_{0}$, is of the order of 0.01 eV (typical for a
high energy phonon).

This mobility behavior together with an increase of the effective
mass is known as large polaron.~\cite{venticinco,ventiseis} Taking
into account that the measured activation energy for the
conductivity is 0.68 eV and using  Eq. 5, the activation energy
for carrier ionization is estimated to be 0.67 eV,  which is
still far from 2.56 eV. Therefore we conclude that the standard
semiconducting mechanism does not satisfactorily explain the
experimental results.

The third mechanism is small polaron
motion.~\cite{ventitres,venticuatro} In some materials where the
interaction between carriers and lattice vibrations is
sufficiently strong, an excess carrier will find it energetically
favorable to remain localized in one of an infinite number of
equivalent sites of the crystal. The carrier-lattice interaction
induces a lattice distortion in the immediate vicinity of the
carrier. The potential well produced by this distortion in turn
acts as a trapping center for the carrier; this self-trapped unit
is called small polaron.~\cite{ventitres,venticuatro}

At temperatures above approximately one half of the Debye
temperature, small polaron motion occurs by thermally activated
hopping. Also, polarons have an associated broad optical
absorption band at photon energies four times larger than the
hopping activation energy.~\cite{ventitres,venticuatro} In our
case, this represents 2.72 eV, which fits quite well with our
observations of 2.56 eV. The small discrepancy can be due to the
electrical field caused by the substitutional {\mgi} ions (which
are negatively charged with respect to the lattice) on the
polaron.

In addition to this phenomenological approach to small polarons,
atomistic simulations of self-trapping of holes in {\alf} have
recently been carried out.~\cite{catorce} These calculations lead
to the conclusion that self-trapped holes are energetically more
favorable than free holes, and that the optical absorption energy
of the self-trapped hole is 2.9 eV. The energies required for the
self-trapped polaron to jump between equivalent positions through
the lattice were also determined. In Fig.15, two types of oxygen
triangles are shown: the so-called small triangles (1, 2, 3 and
4) and a big triangle (5). Triangle 1 lies 2.17 \AA\ above
triangles 2,3,4 and 5; the self-trapped hole is shared by two of
the oxygen atoms in a small triangle; hops inside the small
triangle (involving $60^{o}$ reorientation) require the same
activation energy ($\approx$ 0.9 eV) as hops from a small
triangle (such as 3 in Fig. 15) to another small triangle lying
below or above it such as 1. Any other type of hop requires a
much higher activation energy. Within this picture, a
displacement along the {\it c}-axis requires one hop (and is 2.17
\AA\ long), while a displacement to an equivalent position along a
direction perpendicular to the {\it c}-axis requires four steps:
one hop to a triangle lying in another plane, a reorientation in
this triangle, another hop back to the same plane and a final
reorientation (here, the displacement along the direction of
motion is 4.76 \AA). Although the hop activation energy of 0.9 eV
is not inconsistent with our value (0.68 eV),~\cite{ventisiete}
the anisotropy in the conductivity yields a value of two instead
of four (which is our result). This discrepancy can only be
explained in terms of the frequencies of the phonons involved in
the reorientations and in the hops between triangles lying in
different planes, since the attempt frequency of each hop is
proportional to the phonon frequency. In the above description
the activation energies for the conductivity along the directions
parallel and perpendicular to the {\it c}-axis are the same as
were found in our experiments.

Hall effect measurements could clarify the type of carrier
involved in the electrical conduction. However, we were unable to
perform these measurements because of the high impedance of the
samples, and probably because of their blocking contacts. It was
not possible to measure photocurrents, which could be helpful to
determine the drift mobility of carriers and provide a definite
answer as to whether holes are self-trapped or not.

\figquince

\section*{V. SUMMARY AND CONCLUSIONS}

Optical absorption measurements at 2.56 eV were used to monitor
the concentration of {\mg} centers in two crystals  of {\alf}
intentionally doped with different concentrations of magnesium.
While the Mg concentration was higher in one crystal, the
unintended levels of impurities were also higher. The net effect
was a diminished {\mg} concentration because of charge
compensation. {\mg} centers are formed at temperatures above 1050
K in an oxidizing atmosphere. The increase in concentration with
temperature depends dramatically on the thermal history of the
sample. For samples pre-oxidized at high temperatures, {\mg}
centers form very rapidly and their concentration saturates at
$\approx$ 1500 K. For a non pre-oxidized sample, the saturation
value is achieved after an anneal at 1923 K.

DC and AC electrical measurements were performed to investigate
the electrical conductivity of {\alb} samples with different
concentrations of {\mg} centers in the temperature interval
250-800 K.

At low electrical fields, DC measurements reveal blocking
contacts. At high fields, the I-V characteristic is similar to
that of a diode (corresponding to a blocking contact at one side
of the sample and an ohmic contact at the other side) connected
in series with  the bulk resistance of the sample. The
non-homogeneity of the contacts is responsible for the asymmetry
of the I-V curve with respect to voltage polarity. In addition,
when a current in excess of $\approx$10 {\mA} passes through the
sample, steady EL emission has been observed to emerge from the
negative electrode,~\cite{diecinueve} indicating that the
majority carriers are holes. The intensity of the light is
proportional to current intensity and voltage, showing that ohmic
conduction occurs. Low voltage AC measure
ments show that the
equivalent circuit for the sample is a series combination of the
bulk resistance, {\Rs}, and junction capacitance, {\Cs},
(representing the blocking contacts) connected in parallel with a
capacitance, {\Cp}, which represents the dielectric constant of
the sample. The values determined for the bulk resistance in both
DC and AC experiments are in good agreement. In the temperature
range 273-473 K, {\Rs} diminishes as temperature increases,
whereas {\Cs} and {\Cp} remain practically constant. The
electrical conductivity of {\alb} crystals increases linearly
with the concentration of {\mg} centers. Other impurities serve
as charge compensators to prevent the formation of {\mg} centers.
The conductivity is anisotropic, with the parallel direction four
times higher than in the perpendicular directions to the {\it
c}-axis. The thermal activation energy is 0.68 eV, which is
independent of: 1) the {\mg} content, 2) the crystallographic
orientation, and 3)  the concentration of other impurities.

Neither {\it impurity concentration}~\cite{vente,ventiuno} nor
{\it standard semiconducting}~\cite{ventidos} mechanisms can
adequately explain the observed thermally-activated behavior of
the conductivity. A {\it small-polaron-motion}
mechanism~\cite{ventitres,venticuatro} provides a more likely
explanation. In this mechanism, the interaction between an excess
carrier and the lattice induces a lattice distortion in the
immediate vicinity of the carrier.~\cite{ventitres,venticuatro}
The potential well produced by this distortion traps the carrier,
thereby creating a small polaron.~\cite{ventitres,venticuatro} Our
experimental findings are consistent with the predictions of the
small-polaron-motion mechanism, with the exception that the
experimental value of four is larger than the theoretical
anisotropy of two. This discrepancy is associated with the
frequencies of the phonons involved in the self-trapped polaron
motion.

\section*{Acknowledgments}

Research at the University Carlos III was supported by the CICYT
of Spain. The research of Y.C. is an outgrowth of past
investigations performed at the Solid State Division of the Oak
Ridge National Laboratory.

\end{document}